\documentclass[conference]{IEEEtran}
\IEEEoverridecommandlockouts
\usepackage{cite}
\usepackage{amsmath,amssymb,amsfonts}
\usepackage{algorithmic}
\usepackage{graphicx}
\usepackage{siunitx}
\usepackage{textcomp}
\usepackage{comment}
\usepackage{xcolor}
\usepackage{subcaption}
\usepackage{amsmath}
\usepackage{amssymb}
\usepackage{gensymb}
\usepackage{multirow}
\usepackage{array}
\usepackage{bm}
\usepackage{amsmath}
\usepackage{graphicx}
\usepackage[T1]{fontenc}
\usepackage{lmodern}   % better font support

\usepackage{url}
\usepackage[left=1.62cm,right=1.62cm,top=1.9cm]{geometry}
\usepackage[nolist,nohyperlinks]{acronym}
\usepackage{array}
\usepackage{lipsum}
\usepackage{booktabs}
\usepackage[table]{xcolor}
\usepackage{balance}

\def\BibTeX{{\rm B\kern-.05em{\sc i\kern-.025em b}\kern-.08em
    T\kern-.1667em\lower.7ex\hbox{E}\kern-.125emX}}
\begin{document}

\title{PPO-Based Dynamic Positioning of HAPS-BS in Wind-Disturbed Stratospheric Maritime Networks }

% \author{
% \IEEEauthorblockN {Azim Akhtarshenas$^1$; German Svistunov$^1$; David L\'opez-P\'erez$^{1,2}$, Matteo Bernab\'e$^1$}
% \IEEEauthorblockA{\textit{$^1$Universitat Politècnica de València, Valencia, Spain}} 
% \IEEEauthorblockA{\textit{$^2$Beihang Valencia Polytechnic Institute (BVPI), Hangzhou, China}}
% \textit{aakhtar@upv.edu.es}}

\author{
\IEEEauthorblockN{
Azim Akhtarshenas\IEEEauthorrefmark{1},
German Svistunov\IEEEauthorrefmark{1},
Matteo Bernabè\IEEEauthorrefmark{1},
Kuangyu Zheng\IEEEauthorrefmark{3},
and
David López-Pérez\IEEEauthorrefmark{2}
}
\normalsize\IEEEauthorblockA{\emph{
\IEEEauthorrefmark{1}Universitat Politècnica de València (UPV), Spain} 
}
\normalsize\IEEEauthorblockA{\emph{
\IEEEauthorrefmark{2}Beihang Valencia Polytechnic Institute (BVPI), China}
}
\normalsize\IEEEauthorblockA{\emph{
\IEEEauthorrefmark{3}Beihang University, China} 
}
% \IEEEauthorblockN{Azim Akhtarshenas$^1$; German Svistunov$^1$; David L\'opez-P\'erez$^{1,2}$}
% \IEEEauthorblockA{\textit{$^1$Universitat Politècnica de València (UPV), Valencia, Spain}\\
% $^2$\textit{Beihang Valencia Polytechnic Institute (BVPI), Hangzhou, China}} 
%\IEEEauthorblockA{\textit{$3$}} 
\textit{aakhtar@doctor.upv.es}
% \author{
% \IEEEauthorblockN{
% Azim Akhtarshenas\IEEEauthorrefmark{1}
% German Svistunov\IEEEauthorrefmark{1},
% and
% David López-Pérez\IEEEauthorrefmark{1}\IEEEauthorrefmark{2}
% }
% \normalsize\IEEEauthorblockA{\emph{
% \IEEEauthorrefmark{1}Universitat Politècnica de València (UPV), Spain} 
% }
% \normalsize\IEEEauthorblockA{\emph{
% \IEEEauthorrefmark{2}Beihang Valencia Polytechnic Institute (BVPI), China} 

% }
% \textit{aakhtar@upv.edu.es}
% 
\thanks{
This research is supported by the Generalitat Valenciana, Spain, through the CIDEGENT PlaGenT, Grant CIDEXG/2022/17, Project iTENTE, and the action CNS2023-144333, financed by MCIN/AEI/10.13039/501100011033 and the European Union “NextGenerationEU”/PRTR; and the Xiongan S\&T Innovation Program No.SQ2023XAGG0188 and S\&T Program of Hebei China No.246Z0303G.} 
}

\maketitle

\begin{acronym}[AAAAAAAAAAAAAAAAAAAAAAAA]  % longest acronym to fix width
 \acro{3GPP}{third generation partnership project}
 \acro{4G}{fourth-generation}
 \acro{5G}{fifth-generation}
 \acro{6G}{sixth-generation}
 \acro{AI}{artificial intelligence}
 \acro{BS}{base station}
 \acro{dB}{decibel}
 \acro{dBi}{decibel isotropic}
 \acro{DRL}{deep reinforcement learning}
 \acro{FR}{Frequency Range}
 \acro{FSO}{free-space optical}
 \acro{HAPS}{high-altitude platform station}
 \acro{HAPS-BS}{high-altitude platform station-mounted base station}
 %\acro{HS}{hotspot}
 \acro{IoT}{Internet of things}
 \acro{IoS}{Internet-of-ships}
 \acro{LEO}{low Earth orbit}
 \acro{LoS}{line-of-sight}
 \acro{MCN}{maritime communication network}
 \acro{ML}{machine learning}
 \acro{NLoS}{non-line-of-sight}
 \acro{NTN}{non-terrestrial network}
 \acro{PPO}{proximal policy optimization}
 \acro{PRB}{physical resource block}
 \acro{RL}{reinforcement learning}
 \acro{RSRP}{reference signal received power}
 \acro{SAGIN}{space-air–ground-sea integrated network}
 \acro{SINR}{signal-to-interference-plus-noise ratio}
 \acro{TBS}{terrestrial base station}
 \acro{TN}{terrestrial network}
 \acro{UAV}{unmanned aerial vehicle}
 \acro{UE}{user equipment}
\end{acronym}

\begin{abstract}
%\footnotetext{This research is supported by the Generalitat Valenciana through the CIDEGENT PlaGenT, Grant CIDEXG/2022/17, Project iTENTE, and the action CNS2023-144333, financed by MCIN/AEI/10.13039/501100011033 and the European Union “NextGenerationEU”/PRTR.}
% High-Altitude Platform Stations (HAPS) provide a promising solution for delivering wide-area wireless coverage in maritime regions where terrestrial infrastructure is unavailable. However, maintaining reliable communication performance is challenging due to dynamic ship mobility and atmospheric disturbances, particularly stratospheric wind turbulence that affects HAPS positioning.
% In this paper, we propose a deep reinforcement learning (DRL)-based framework for dynamic positioning of wind-disturbed HAPS base stations in maritime communication networks. In the proposed architecture, a centralized DRL agent deployed on a coordinator HAPS controls the navigation of multiple serving HAPS based on radio measurements and network feedback, capturing realistic channel conditions and user mobility patterns. A Proximal Policy Optimization (PPO) algorithm is employed to learn robust positioning policies that improve coverage stability and system throughput under wind-induced disturbances.
% Simulation results demonstrate that the proposed DRL-based approach effectively mitigates wind-induced positioning deviations while maintaining reliable wide-area connectivity for maritime users.

High-Altitude Platform Stations (HAPS) offer a promising solution for wide-area wireless coverage in maritime regions lacking terrestrial infrastructure. However, maintaining reliable performance is challenging due to dynamic ship mobility and atmospheric disturbances, particularly stratospheric wind effects on HAPS positioning.
This paper proposes a deep reinforcement learning (DRL)-based framework for dynamic positioning of wind-disturbed HAPS-mounted base stations in maritime networks. A centralized DRL agent deployed on a coordinator HAPS controls multiple serving HAPS using radio measurements and network feedback, capturing realistic channel conditions and user mobility. A Proximal Policy Optimization (PPO) algorithm is employed to learn robust positioning policies that enhance coverage stability and system throughput under wind disturbances.
Simulation results show that the proposed approach effectively mitigates wind-induced positioning deviations while ensuring reliable wide-area connectivity for maritime users.

Index Terms - Non-Terrestrial Networks, HAPS, Maritime Communication Networks, DRL, PPO, Wind-Aware Positioning
\end{abstract}

\section{Introduction}

The increasing demand for ocean-based services, including shipping logistics, offshore monitoring, tourism, fishing, and emergency response, has intensified the need for reliable maritime communication networks (MCNs) \cite{alqurashi2022maritime, shang2025maritime}. However, providing continuous wireless connectivity over open-sea regions remains challenging due to the lack of terrestrial infrastructure, large coverage requirements, harsh propagation environments, and vessel mobility \cite{niknami2023maritime}. Conventional terrestrial networks are mainly effective near coastlines, while satellite-based systems provide global coverage at the cost of higher latency, limited per-user capacity, and frequent handovers \cite{shang2025maritime}.

To overcome these limitations, non-terrestrial networks (NTNs) have been widely investigated as a key enabler for future maritime connectivity. In particular, space-air-ground-sea integrated networks combine terrestrial stations, satellites, aerial platforms, and maritime users to provide flexible and seamless coverage over oceanic regions \cite{wei2021hybrid}. Within this architecture, \acp{HAPS} have emerged as a promising intermediate layer between satellites and \acp{UAV}. Operating in the stratosphere at approximately $20\,\mathrm{km}$ altitude, HAPS can provide wide-area coverage, lower propagation delay than satellites, and longer endurance than UAVs, making them ideal candidates for persistent MCN services 
\cite{svistunov2025bridging}.
% \cite{svistunov2025bridging, elkhazraji2025haps}.
Recent studies have investigated HAPS as aerial base stations for coverage extension, topology optimization, and maritime connectivity enhancement. HAPS-mounted base stations (HAPS-BSs) can support large service areas and provide flexible deployment for offshore users \cite{duan2017optimal, cao2020topological, lin2026haps}. However, maintaining reliable communication performance requires accurate positioning of HAPS-BSs, especially when directional antennas are used and maritime users are mobile. In many existing works, HAPS locations are assumed to be static or quasi-static, which limits their applicability in realistic maritime environments \cite{alam2021high}.

A key challenge for practical HAPS deployment is the presence of stratospheric wind. Unlike terrestrial base stations, HAPS are affected by horizontal wind flows, temporal wind variability, and altitude-dependent atmospheric dynamics \cite{cui2026reinforcement, delgado2024station}. Operational data from stratospheric balloon systems, such as Project Loon, show that zonal and meridional winds vary over time and can introduce residual variability of approximately $3\,\mathrm{m/s}$ \cite{friedrich2017comparison}. Such wind-induced displacement can move HAPS away from their desired coordinates, degrade antenna alignment, reduce link quality, and impact coverage stability.
Several works have addressed HAPS station-keeping and wind-aware trajectory planning by exploiting altitude-dependent wind layers or optimizing platform motion under dynamic wind fields \cite{du2019station, hirai2025maritime}. However, these approaches often require explicit wind-field knowledge or predefined trajectory rules, which are difficult to apply in highly dynamic maritime environments where user mobility must also be considered.

To address this challenge, machine learning (ML) approaches, particularly deep reinforcement learning (DRL), provide a suitable framework for continuous control under uncertainty \cite{saafi2022ai, xu2025deep}. In particular, proximal policy optimization (PPO) enables stable learning in continuous action spaces and has been successfully applied to aerial platform control and resource management \cite{ibanez2025optimizing}. Instead of relying on detailed aerodynamic modeling, the HAPS positioning problem can be formulated as a mobility-control task in which the agent learns to follow maritime hotspots while compensating for wind-induced deviations.

Compared to our previous work~\cite{ibanez2025optimizing}, where PPO was applied for UAV-mounted base station (UAV-BS) positioning operating at $50\,\mathrm{m}$, this paper considers HAPS-BSs at $20\,\mathrm{km}$, where weaker signals make PPO observation design more challenging. Unlike the UAV-BS setup with omnidirectional antennas, HAPS-BSs employ directional reflector antennas, requiring accurate alignment for reliable signal reception.
Furthermore, instead of a simple throughput-based reward, we adopt a sigmoid-based formulation with systematically tuned parameters, improving training stability and convergence.
Motivated by the limitations of existing maritime communication solutions in wind-affected environments and our prior work~\cite{ibanez2025optimizing} on platform positioning, we investigate a DRL-based HAPS-BS framework for dynamic maritime networks. The model captures moving hotspots and stratospheric wind dynamics, enabling adaptive positioning under user mobility and atmospheric perturbations to ensure reliable coverage and improved throughput. 
% Motivated by our previous study~\cite{ibanez2025optimizing}, the wide-area coverage of \ac{HAPS-BS} and the limitations of terrestrial, satellite, and \ac{UAV}-based maritime communication solutions, this paper investigates an intelligent HAPS-BS-assisted \ac{MCN}, where \ac{HAPS-BS} are considered the primary connectivity provider. The proposed model considers moving maritime hotspots and a wind-affected stratospheric environment, where HAPS motion is influenced by mean horizontal wind, smooth temporal variation, and correlated residual disturbance. The objective is to enable HAPS-BSs to maintain reliable coverage and improve throughput while adapting to both user mobility and atmospheric perturbations.
The main contributions of our paper are summarized as follows:
\begin{itemize}
\item We model flexible and continuous HAPS mobility to capture realistic trajectory dynamics in maritime environments using refined reward function. 
\item We leverage reference signal measurements from UEs to enable intelligent HAPS-BS control without requiring explicit location information.
\item We model stratospheric wind dynamics, including mean flow, temporal variation, and stochastic disturbances, and incorporate them into the learning environment for robust HAPS-BS positioning.
\end{itemize}
\section{Stratospheric Wind Model}

In this work, the HAPS-BS operates at an altitude of approximately 20 km, corresponding to the lower stratosphere. At this altitude, wind behavior is fundamentally different from low-altitude atmospheric conditions, where gusts and strong turbulence dominate. Instead, the wind field is primarily characterized by large-scale horizontal flow with relatively smooth temporal evolution and moderate stochastic variability.
Following observations from stratospheric balloon experiments, particularly the Loon project~\cite{friedrich2017comparison}, the wind at this altitude can be modeled as a combination of three components:
\begin{equation}
\mathbf{v}_{\text{wind}}(t) = \mathbf{v}_{m} + \mathbf{v}_{s}(t) + \mathbf{v}_{\text{res}}(t),
\end{equation}
where $\mathbf{v}_{m}$ represents the dominant horizontal wind, $\mathbf{v}_{s}(t)$ captures slow temporal variations, and $\mathbf{v}_{\text{res}}(t)$ models stochastic residual fluctuations.
Empirical results indicate that the residual variability follows an approximately Gaussian distribution with a standard deviation in the range of $2.5$--$3.5$ m/s (Figure 2.~\cite{friedrich2017comparison}). The temporal evolution of the wind is smooth and correlated over time, which can be effectively modeled using an autoregressive process (AR).
The dominant wind direction and magnitude are influenced by large-scale atmospheric circulation patterns, such as jet streams and stratospheric layers, as discussed in \cite{coy2017global, xu2022station}. These studies highlight that, at stratospheric altitudes, wind fields are better represented as continuous and slowly varying processes rather than discrete gust events typical of lower altitudes.
Based on these observations, the wind model used in our study incorporates:
(i) a constant mean horizontal component,
(ii) a slow sinusoidal variation to capture large-scale changes, and
(iii) a correlated stochastic component to represent residual variability.
This formulation provides a realistic yet tractable representation of stratospheric wind conditions for RL-based HAPS control. For more information and details, refer to~\cite{friedrich2017comparison,coy2017global,xu2022station}.
\section{System Model}
\label{sec:system_model}
In this section, we describe the proposed system model. We consider the downlink performance of a non-terrestrial HAPS-based \ac{MCN} serving multiple mobile maritime UEs. 
The system operates in both downlink and uplink using TDMA.
Let $\mathcal{U} = \{u_1, \ldots, u_U\}$ denote the set of \acp{UE}, with positions $\boldsymbol{{\rho}}^\mathrm{U}_u = [x^\mathrm{U}_u, y^\mathrm{U}_u, z^\mathrm{U}_u]^\top$. The collective \acp{UE} positions are given by $\boldsymbol{\mathrm{X}}^{\mathrm{U}} = [\boldsymbol{{\rho}}^{\mathrm{U}}_1, \ldots, \boldsymbol{{\rho}}^{\mathrm{U}}_U]$. UEs are grouped into hotspots, where the communication metrics of all UEs within each hotspot are aggregated, averaged, and, when required, represented by a single equivalent UE.
Let $\mathcal{H} = \{h_1, \ldots, h_H\}$ denote the set of hotspots, each with identical radius $r$ and an equal number of UEs, i.e., $N = U/H$ per hotspot. The hotspot mobility follows a linear model, moving from left to right and vice versa, with all UEs traveling at a constant speed $v$.
Network coverage is provided by a NTN consisting of $D$ HAPS-BSs operating at an altitude of $20\,\mathrm{km}$. Each hotspot is associated with a dedicated HAPS-BS.
Let $\mathcal{D} = \{d_1, \ldots, d_D\}$ denote the set of HAPS-BSs, with positions $\boldsymbol{{\rho}}^{\mathrm{D}}_d = [x^{\mathrm{D}}_d, y^{\mathrm{D}}_d, z^{\mathrm{D}}_d]^\top$. The collective HAPS-BS positions are given by $\boldsymbol{\mathrm{X}}^{\mathrm{D}} = [\boldsymbol{{\rho}}^{\mathrm{D}}_1, \ldots, \boldsymbol{{\rho}}^{\mathrm{D}}_D]$.

\begin{figure}[t]
    \centering    \includegraphics[width=.8\linewidth,height=7.5cm,keepaspectratio]{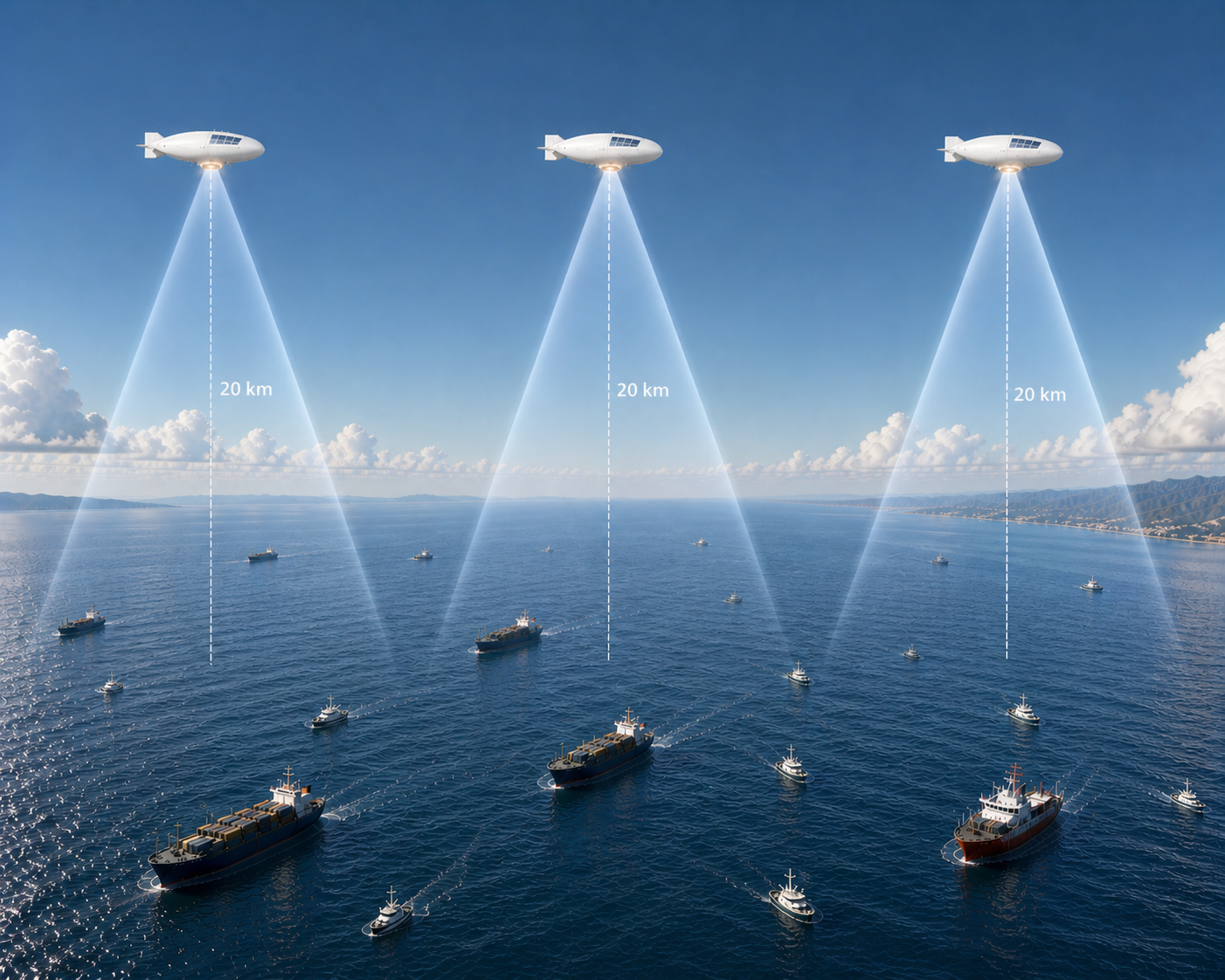}
    \caption{HAPS-assisted maritime communication.}
    \label{fig:HAPS_OCEAN}
\end{figure}

%[German]: I didn't find any details about the role of the "central controller HAPS" in the paper. Is it a real part of simulation scenario? If no, I believe it's better to remove it from the picture and never mention.

% Let $\mathcal{H} = \{1, \ldots, h, \ldots, H\}$ denote the set of serving HAPS-BSs, where $H = |\mathcal{H}|$ is the total number of HAPS-BSs (The blue ones in~Fig.~\ref{fig:HAPS_OCEAN}). 
% The three-dimensional position of the $h$-th HAPS-BS is represented by $\boldsymbol{\rho}^{\mathrm{H}}_h = [x^{\mathrm{H}}_h, y^{\mathrm{H}}_h, z^{\mathrm{H}}_h]^{\top}$, where $x$, $y$, and $z$ denote the Cartesian coordinates in three-dimensional space. 
% Similarly, let $\boldsymbol{\rho} = [x^{\mathrm{C}}, y^{\mathrm{C}}, z^{\mathrm{C}}]^{\top}$ represent the three-dimensional position of the central controller HAPS (The black one in~Fig.~\ref{fig:HAPS_OCEAN}).
% In this study, the central \ac{HAPS}is assumed to be located at the geometric center of the considered area to ensure balanced proximity to all serving HAPS-BSs.
% The collective positions of all HAP-BSs are organized in the matrix $\boldsymbol{\rho}^{\mathrm{H}} = [\boldsymbol{\rho}^{\mathrm{H}}_1, \ldots, \boldsymbol{\rho}^{\mathrm{H}}_H]$.

\subsubsection{Channel Model}
Each HAPS-BS is equipped with a circular-aperture reflector antenna, as specified in~\cite{3GPP38811}, with the reflector fully downtilted. 
%Thus, each \ac{HAPS-BS} forms a dynamic coverage cell determined by the corresponding \ac{HAPS} position.
For the $u^\text{th}$ \ac{UE}, served by $d^\text{th}$ HAPS-BS, the large-scale channel gain $\beta_{u,d}$ is defined as, $    \beta_{u,d} = p_{u,d}\,\tau_{u,d}\, g^r_{u,d}\,\,, $ 
% \begin{equation}
%     \beta_{u,c} = p_{u_c}\,\tau_{u_c}\, g^r_{u,c}\,\,, 
% \end{equation}
where $p_{u,d}$ is the path loss gain,  $\tau_{u,d}$ is the shadowing gain and $g_{u,d}$ is the reflector antenna gain, computed from the \ac{3GPP} statistical channel model defined  in~\cite{3GPP38811}.
Specifically, path loss gain $p_{u,d}$ is modeled following ITU-R recommendations~\cite{ITU-R_P.618-14}  as follows, 
\begin{equation}
        p_{u,d} = 1/{p_{u,d}^{\rm fspl} \, p_{u,d}^{\rm cl} \, p_{u,d}^{\rm ga} \, p_{u,d}^{\rm ra} \, p_{u,d}^{\rm ca} \, p_{u,d}^{\rm sa}}\,,
\end{equation}
where $p_{u,d}^{\rm fspl}$ is the free space propagation loss, $p_{u,d}^{\rm cl}$ is the clutter loss and $p_{u,d}^{\rm ga}$ $p_{u,d}^{\rm ra}$, $p_{u,d}^{\rm ca}$ $p_{u,d}^{\rm sa}$ are the gaseous, rain, cloud, and scintillation attenuation, respectively.
The small-scale channel ${\bf h}_{u,d,k}$ for each frequency resource \(k\), is modelled as a Rician fading channel and computed as,
\begin{equation}\label{eq:ComplexChannelRician}
    {\bf h}_{u,d,k} =
    \sqrt{\frac{K}{1+K}} \, {\bf h}^{\rm LOS}_{u,d,k}
    +
    \sqrt{\frac{1}{1+K}} \,  {\bf h}^{\rm NLOS}_{u,d,k} ,
\end{equation}
where $K$ denotes the Rician factor, ${\bf h}^{\rm NLOS}_{u,d,k}$ represents the \ac{NLoS} channel component modeled as Rayleigh fading, and ${\bf h}^{\rm LOS}_{u,d,k}$ represents the \ac{LoS} component which follows a plane-wave approximation as defined in~\cite{3GPP38901}.

\subsubsection{User association, SINR and throughput computation}
Each UE within a hotspot is associated with the HAPS-BS dedicated to a corresponding hotspot. %For a given \ac{UE} \(u\) and candidate cell \(c\), the \ac{RSRP} is computed as the average received power of the corresponding reference signals, accounting for transmit power, antenna gains, and the aforementioned channel model effects.
The downlink \ac{SINR} for the $u^{\text{th}}$ UE, served by the $\hat{d}^{\text{th}}$ HAPS-BS, is computed as:
\begin{equation}
    \gamma_{u,\hat{d},k}(\bm{\rho}^{\mathrm{U}}_{u},\bm{\mathrm{{X}}}^{D}) =
    \frac{
    \beta_{u,\hat{d}}
        \left| h_{u,\hat{d},k} \right|^2
        w_{u,\hat{d},k}
    }{
        \sum\limits_{d \neq \hat{d}}
        \sum\limits_{u' \in \mathcal{U}_d}
        \beta_{u,d}
        \left| h_{u,d,k} \right|^2
        w_{u',d,k}
        + \sigma_k^2
    }
    \label{eq:SINR_Computation_NTN}
\end{equation}
where 
\(\mathcal{U}_d\) denotes the set of UEs served by the $d^\text{th}$ (not $\hat{d}^\text{th}$) HAPS-BS,
\(w_{u,d,k}\)~denote the transmit power allocated by the $d^\text{th}$ HAPS-BS to the $u^\text{th}$ UE on the $k^\text{th}$ frequency resource,
\(\sigma_k^2\) represents the noise power on the $k^\text{th}$ frequency resource. 
In this formulation, the reflector is modeled as a single antenna, and therefore the channel vector $\mathbf{h}_{u,d,k}$ reduces to a scalar ${h}_{u,d,k}$ and no antenna precoding is applied.
The effective \ac{SINR} of the $u^\text{th}$ UE from the $h^\text{th}$ hotspot and associated with the $d^\text{th}$ HAPS-BS,
denoted by $\tilde{\gamma}_{u,h,{d}}$, 
is then obtained from the set of per frequency resource \acp{SINR} \(\gamma_{u,{d},k}\), using a mutual-information-based effective-\ac{SINR} mapping framework. 
This effective \ac{SINR} provides an overall representation of the quality experienced by the \ac{UE} over its allocated frequency resources.

Assuming round-robin scheduling and full-buffer traffic,
the achievable downlink throughput of the $u^\text{th}$ \ac{UE} from the \(h^{\text{th}}\) hotspot associated with the $d^\text{th}$ HAPS-BS, with positions \(\bm{\rho}^{\mathrm{U}}\) and \(\bm{\rho}^{\mathrm{D}}\), respectively, is computed as
\begin{equation}
    R_{u,h,{d}} (\bm{\rho}^{\mathrm{U}}_{u},\bm{\mathrm{{X}}}^{D}) = 
    \frac{
    N^{\rm FR}_{{d}} \, B^{\rm FR}_{{d}}
    }{
    N^{\rm UE}_{{d}}
    }
    \log_2\!\left(1+\tilde{\gamma}_{u,h,{d}}(\bm{\rho}^{\mathrm{U}}_{u},\bm{\mathrm{{X}}}^{D})\right),
    \label{eq:AchievableDataRate}
\end{equation}
where \(N^{\rm FR}_{{d}}\) is the total number of frequency resource blocks available at the serving HAPS-BS \({d}\), 
\(B^{\rm FR}_{{d}}\) is the bandwidth of each frequency resource, 
and \(N^{\rm UE}_{{d}}\) is the number of UEs scheduled to the serving HAPS-BS \({d}\).

\subsubsection{Objective function}
The objective is to determine in real time the HAPS-BS positions $\bm{\mathrm{X}}^{\mathrm{D}}$ that maximizes the total fair throughput. 
This metric, for a given $\bm{\mathrm{X}}^{\mathrm{U}}$ and $\bm{\mathrm{X}}^{\mathrm{D}}$, is defined as the sum of logarithmic UE throughputs~\cite{9878252}:
\begin{equation}
R_{\mathrm{fair}}(\bm{\mathrm{X}}^{\mathrm{U}}, \bm{\mathrm{X}}^{\mathrm{D}})
=
\sum_{d=1}^{D}
\sum_{u \in \mathcal{U}_d}
\log_{10}
R_{u,h,d}\!\left(\bm{\rho}^{\mathrm{U}}_{u}, \bm{\mathrm{X}}^{\mathrm{D}}\right)
\label{eq_intro:fair_rate}
\end{equation}
where $\mathcal{U}_d$ denotes the set of UEs associated with the $d$-th HAPS-BS. 
This formulation promotes fairness by prioritizing improvements for low-throughput UEs.
The optimization problem is thus:
\begin{equation}
\underset{\bm{\mathrm{X}}^\mathrm{D}}{\text{max}}
R_{\mathrm{fair}}(\bm{\mathrm{X}}^\mathrm{U}, \bm{\mathrm{X}}^\mathrm{D}).
\label{fairfair}
\end{equation}
The problem is challenging due to its high dimensionality, stochasticity, and non-linearity.
The continuous movement of UEs, radio channel variations, and stratospheric wind dynamics require constant adaptation, making conventional methods struggle with system complexity. Given these challenges, ML, particularly RL, emerges as a promising approach \cite{sutton2018reinforcement}. RL enables real-time interaction with the environment to learn robust policies that maximize the total fair throughput $R_{\text{fair}}$ while adapting to UE mobility, channel variations, and wind disturbances.
%}
\section{Proposed RL Solution}
A centralized architecture is adopted, where a single \ac{RL} agent controls the HAPS-BSs under stratospheric wind dynamics, including stochastic wind variations, to provide communication services. 
In this framework, the agent interacts with the environment through states $S$, actions $A$, rewards $\mathcal{R}$, and a policy $\pi$ mapping states to actions. The state captures user mobility, channel variations, and wind-induced perturbations.

\subsection{Proposed Proximal Policy Optimization}
To address the HAPS-BS positioning problem, we employ PPO \cite{schulman2017proximal}, a state-of-the-art policy gradient method suited for continuous control. PPO is an on-policy, model-free actor--critic algorithm that uses the advantage function $\hat{A}_t = r_t - V(s_t)$, where $s_t$ and $r_t$ denote the state and reward at time step $t$, and $V(s_t)$ is the value function. This enables joint optimization of the policy $\pi$ and the value function $V(s)$.
Compared with classical policy gradient methods, PPO improves training stability via a clipped surrogate objective and supports continuous state and action spaces without discretization, making it suitable for wind-affected HAPS control.

\subsubsection{State-Space Definition}
We define the state-space representation for the DRL agent in the HAPS trajectory optimization problem. Each HAPS-BS operates over $T$ episodes, interacting with the environment at each time step under stratospheric wind disturbances. A time-indexed state representation capturing the dynamics of the $d^{\text{th}}$ HAPS-BS and the $u^{\text{th}}$ UE is given by:
\begin{equation}
\tilde{\boldsymbol{{\rho}}}^{\mathrm{D}}_d =
\left[
\boldsymbol{\rho}_{d,t-T+1}^{\mathrm{D}}, \dots, \boldsymbol{\rho}_{d,t-i}^{\mathrm{D}}, \dots, \boldsymbol{\rho}_{d,t}^{\mathrm{D}}
\right],
\end{equation}
\begin{equation}
\tilde{\gamma}_{u,h,d} = % [German]: should we denote this gamma bold also?
\left[
\gamma_{u,h,d,t-T+1}, \dots, \gamma_{u,h,d,t-i}, \dots, \gamma_{u,h,d,t}
\right],
\end{equation}
\begin{equation}
\tilde{\alpha}_{u,h,d} = % [German]: should we denote this alpha bold also?
\left[
\alpha_{u,h,d,t-T+1}, \dots, \alpha_{u,h,d,t-i}, \dots, \alpha_{u,h,d,t}
\right].
\end{equation}

Here, $\{u,h,d,t\}$ denote the $u^{\text{th}}$ UE, $h^{\text{th}}$ hotspot, $d^{\text{th}}$ HAPS-BS, and time step, respectively. $\tilde{\boldsymbol{\rho}}^{\mathrm{D}}_d$ represents the position history of the $d^{\text{th}}$ HAPS-BS, which is affected by stratospheric wind dynamics and turbulence, while $\tilde{\gamma}_{u,h,d}$ and $\tilde{\alpha}_{u,h,d}$ denote the corresponding SINR and angle of arrival (AoA) sequences. The instantaneous state at time step $i$ is defined as
$s_{t=i} =
\bigl\{
\boldsymbol{\rho}^{\mathrm{D}}_{d,i},\,
\gamma_{u,h,d,i},\,
\alpha_{u,h,d,i}
\bigr\}$,
and we use $s_i$ for notational simplicity.
For the network described in Section~\ref{sec:system_model}, consisting of $D$ HAPS-BSs and $H$ hotspots, a one-to-one association is assumed ($d = h$ and $D=H$). The global spatial representation at time step $i$ is
\begin{equation}
\mathbf{X}^{\mathrm{D}}_{i}
=
\begin{bmatrix}
\boldsymbol{\rho}^{\mathrm{D}}_{1,i} \; \cdots \; \boldsymbol{\rho}^{\mathrm{D}}_{d,i} \; \cdots \; \boldsymbol{\rho}^{\mathrm{D}}_{D,i}
\end{bmatrix}.
\end{equation}
The corresponding SINR and AoA matrices for the entire proposed network with $N \times D$ UEs are defined as
\begin{equation}
\boldsymbol{\Gamma}_{i}
=
\begin{bmatrix}
\gamma_{1,1,1,i} & \cdots & \gamma_{u,1,1,i} & \cdots & \gamma_{N,1,1,i} \\
\vdots & \ddots & \vdots & \ddots & \vdots \\
\gamma_{1,h',d',i} & \cdots & \gamma_{u',h',d',i} & \cdots & \gamma_{N',h',d',i} \\
\vdots & \ddots & \vdots & \ddots & \vdots \\
\gamma_{1,H,D,i} & \cdots & \gamma_{u'',H,D,i} & \cdots & \gamma_{N'',H,D,i}
\end{bmatrix},
\end{equation}
\begin{equation}
\boldsymbol{\mathrm{A}}_{i}
=
\begin{bmatrix}
\alpha_{1,1,1,i} & \cdots & \alpha_{u,1,1,i} & \cdots & \alpha_{N,1,1,i} \\
\vdots & \ddots & \vdots & \ddots & \vdots \\
\alpha_{1,h',d',i} & \cdots & \alpha_{u',h',d',i} & \cdots & \alpha_{N',h',d',i} \\
\vdots & \ddots & \vdots & \ddots & \vdots \\
\alpha_{1,H,D,i} & \cdots & \alpha_{u'',H,D,i} & \cdots & \alpha_{N'',H,D,i}
\end{bmatrix}.
\end{equation}

Here, $\gamma_{u',h',d',i}$ represents the SINR of the $u'^{\text{th}}$ UE in the $h'^{\text{th}}$ hotspot served by the $d'^{\text{th}}$ HAPS-BS at time step $i$, while $\alpha_{u',h',d',i}$ denotes the corresponding AoA measurement.
By applying the arithmetic mean to the SINR, and the mean and standard deviation to the AoA across $N$ UEs per hotspot, we obtain the network-level representation as:
\begin{equation}
\boldsymbol{\gamma}_{i} =
\left[
\bar{\gamma}_{1,1,i}, \dots, \bar{\gamma}_{h,d,i}, \dots, \bar{\gamma}_{H,D,i}
\right]^{\top},
\end{equation}
\begin{equation}
\boldsymbol{\psi}_{i} =
\left[
\psi_{1,1,i}, \dots, \psi_{h,d,i}, \dots, \psi_{H,D,i}
\right]^{\top},
\end{equation}
\begin{equation}
\boldsymbol{\Omega}_{i} =
\left[
\Omega_{1,1,i}, \dots, \Omega_{h,d,i}, \dots, \Omega_{H,D,i}
\right]^{\top}.
\end{equation}

Here, $\bar{\gamma}_{h,d,i}$, $\psi_{h,d,i}$, and $\Omega_{h,d,i}$ denote the average SINR, mean AoA, and AoA standard deviation of the $h^{\text{th}}$ hotspot served by the $d^{\text{th}}$ HAPS-BS at time step $i$.  So, for the $d^{\text{th}}$ HAPS-BS serving the $h^{\text{th}}$ hotspot, the RL agent observes the system in an aggregated form. 
% Assuming three UAV-BSs operating at a fixed altitude and serving three hotspots, the overall state space is defined as:
% % --- Reduce equation spacing locally ---
% \begingroup
% \setlength{\abovedisplayskip}{6pt}
% \setlength{\belowdisplayskip}{6pt}

% \begin{equation}
% \label{eq:observation_set}
% \mathcal{\mathbf{S}} =
% \left\{
% \begin{aligned}
% &\big((x_{1,i},y_{1,i}),\, \bar{\gamma}_{1,1,i},\, \Psi_{1,1,i},\, \Phi_{1,1,i}\big),\\
% &\big((x_{2,i},y_{2,i}),\, \bar{\gamma}_{2,2,i},\, \Psi_{2,2,i},\, \Phi_{2,2,i}\big),\\
% &\big((x_{3,i},y_{3,i}),\, \bar{\gamma}_{3,3,i},\, \Psi_{3,3,i},\, \Phi_{3,3,i}\big),
% \end{aligned}
% \right\}.
% \end{equation}
% \endgroup  %Azim done
% The resulting feature vector is
% $\boldsymbol{s}_{d,i} =
% \begin{bmatrix}
% \boldsymbol{\rho}^{\mathrm{D}}_{d,i}, &
% \gamma_{h,d,i}, &
% \psi_{h,d,i}, &
% \Omega_{h,d,i}
% \end{bmatrix}$,
% which is used as input to the neural network.

\begin{figure*}[!t]
    \centering

    \begin{subfigure}[t]{0.484\linewidth}
        \centering
        \includegraphics[width=\linewidth,height=4.8cm,keepaspectratio]{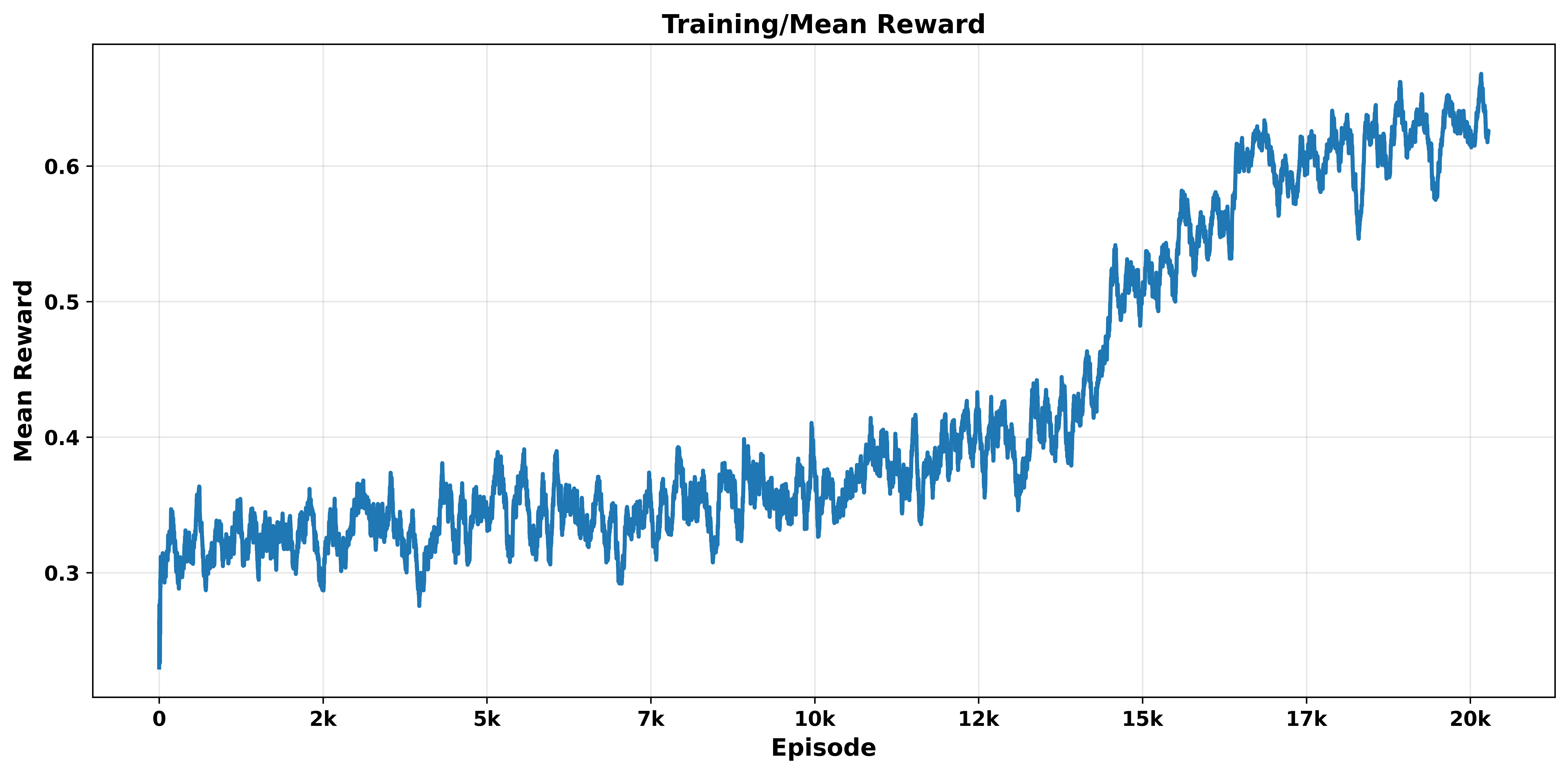}
        \caption{\textbf{ Mean training reward vs. episodes.}}
        \label{fig1:train_mean_reward}
    \end{subfigure}
    \hfill
    \begin{subfigure}[t]{0.484\linewidth}
        \centering
        \includegraphics[width=\linewidth,height=4.8cm,keepaspectratio]{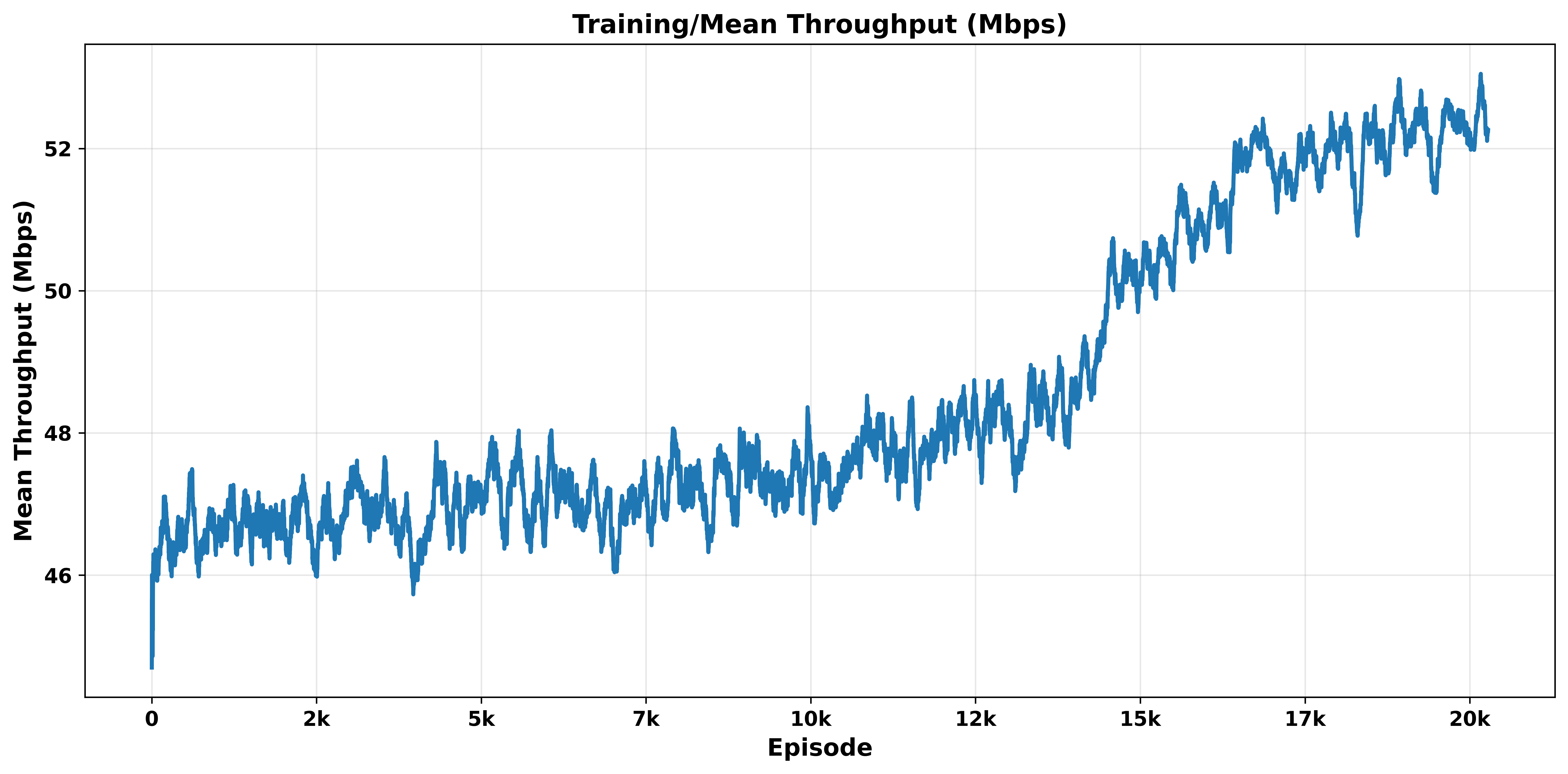}
        \caption{\textbf{Mean training throughput vs. episodes.}}
        \label{fig2:train_mean_throughput}
    \end{subfigure}

    \vspace{0.2em}

    \begin{subfigure}[t]{0.484\linewidth}
        \centering
        \includegraphics[width=\linewidth,height=4.8cm,keepaspectratio]{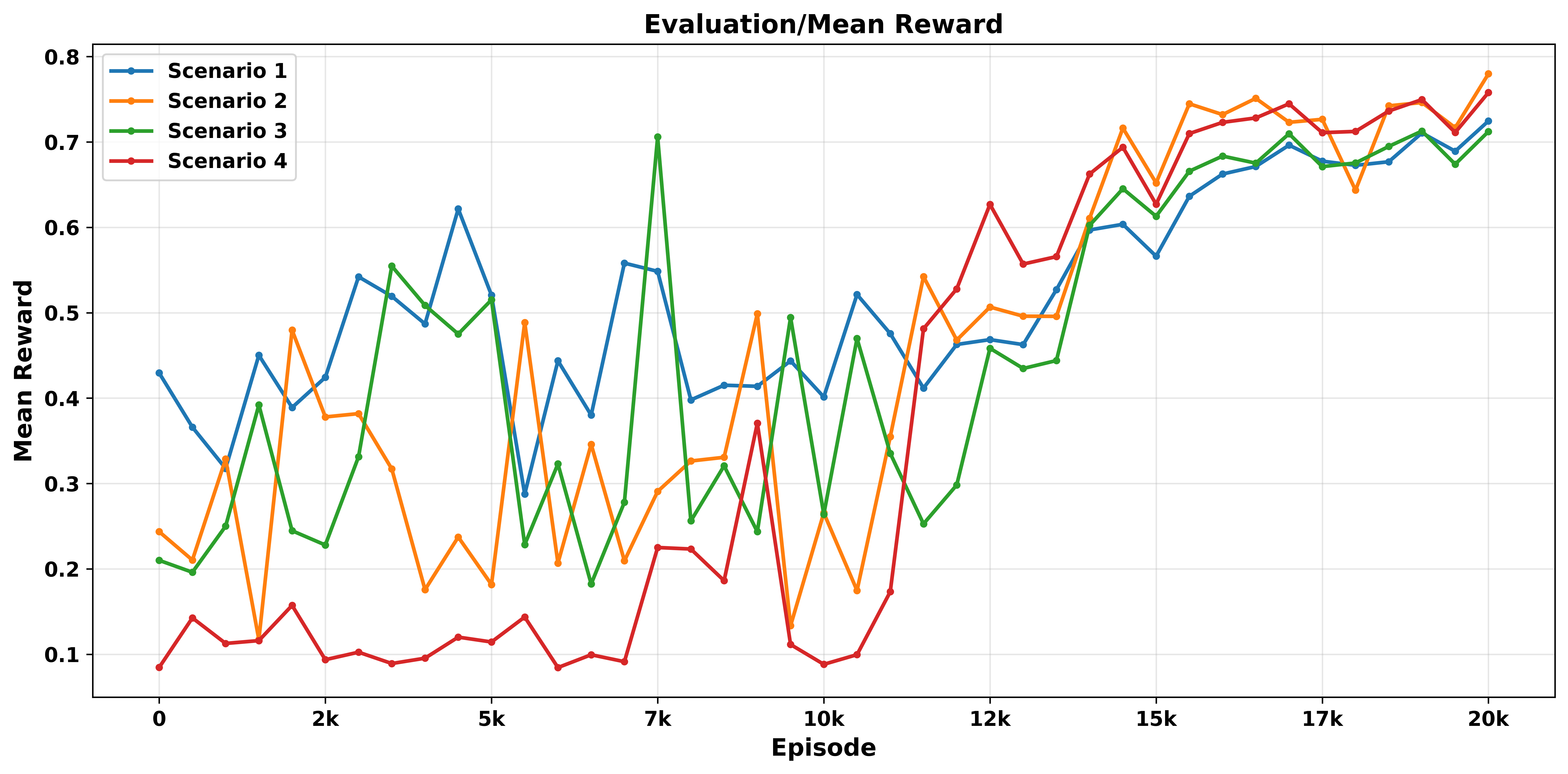}
        \caption{\textbf{Mean evaluation reward vs. episodes.}}
        \label{fig3:eval_mean_reward}
    \end{subfigure}
    \hfill
    \begin{subfigure}[t]{0.484\linewidth}
        \centering
        \includegraphics[width=\linewidth,height=4.8cm,keepaspectratio]{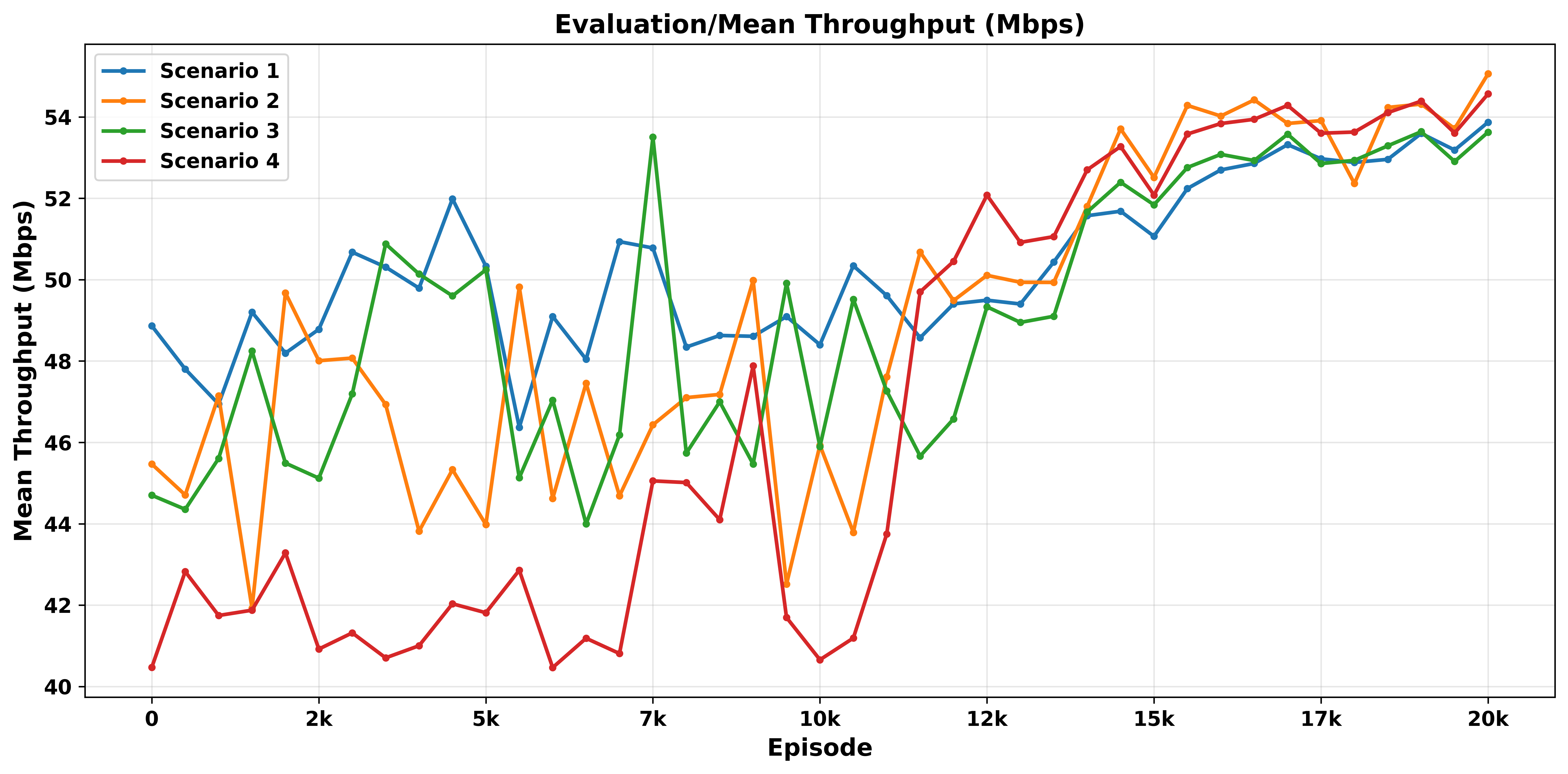}
        \caption{\textbf{Mean evaluation throughput vs. episodes.}}
        \label{fig4:eval_mean_throughput}
    \end{subfigure}

    \caption{Performance of the proposed HAPS-BS network under stratospheric wind disturbances at an altitude of $h = 20\,\mathrm{km}$.}
    \label{fig:four_plots}
\end{figure*}
\subsubsection{Action Space}
The action space is continuous, where the agent selects the movement direction and magnitude, defined by the angle $\alpha_d$ and distance $r$, respectively. The action space is given by $\mathcal{A} = [-180^\circ, 180^\circ) \times [0, r_{\max}]$, where $r_{\max}$ denotes the maximum distance the HAPS-BS can travel in a single action, and $\alpha_d$ is measured with respect to the east.
Our implementation leverages PPO's capability to operate in continuous action spaces. By avoiding discretization, the agent can explore a continuous set of positions, enabling precise adjustments of the HAPS-BS location. This is particularly advantageous, as even small position changes, induced by weak wind variations, can significantly impact network performance.

% \subsubsection{Action space}
% As the action space is continuous, the agent selects two variables: the movement direction and magnitude. The direction is defined by the angle $\alpha_d$, and the magnitude by the distance $r$. Accordingly, the \textbf{action space} is defined as $\mathcal{A} = \{(\alpha_d, r)\}$
% % \begin{equation}
% % \mathcal{A} = \{(\alpha_d, r)\},
% % \end{equation}
% where $\alpha_d \in [-180, 180)$ and $r \in [0, r_{\max}]$. Here, $r_{\max}$ denotes the maximum distance the HAPS-BS can travel in a single action, and $\alpha_d$ is measured with respect to the east.
% Our implementation leverages PPO's capability to operate in continuous action spaces. By avoiding action discretization, the agent can explore a continuous set of positions, allowing precise adjustments of the HAPS-BS location. This is particularly advantageous in our scenario, where small position changes can significantly affect network performance.
%  {\color{blue}where ${\boldsymbol{\gamma}}_{h,d,i}$ denote the vectors of SINR, within the $h^{\text{th}}$ hotspot served by the $d^{\text{th}}$ HAPS-BS.} 
% Furthermore, ${\boldsymbol{\alpha}}_{h,d,i}$ denotes the vector of AoA measurements for the same hotspot, where $\alpha_{u,h,d,i} \in (-\pi,\pi]$ represents the AoA of the $u^{\text{th}}$ UE, for $u \in \{1,\ldots,N\}$ and $i \in \{0,\ldots,T-1\}$.
\subsubsection{Reward}
The reward aims to maximize the fairness-aware throughput $R_{\mathrm{fair}}$ defined in (\ref{eq_intro:fair_rate}). To ensure consistent scaling and stabilize training, we normalize the reward using a sigmoid mapping:
$\mathcal{R} = \left(1 + \exp\!\left(-c_s (R_{\mathrm{fair}} - c_m)\right)\right)^{-1}$,
where $c_s$ controls the slope and $c_m$ sets the midpoint ($\mathcal{R}=0.5$). The parameters are tuned based on the throughput achieved when HAPS-BSs are directly above their corresponding hotspots, as well as the average throughput over entire episodes, measured across multiple realizations.

\section{Numerical Results and Discussion}
\subsubsection*{Scenario}
To evaluate the effectiveness of the proposed algorithm, we consider a $1.5\,\mathrm{km} \times 1.5\,\mathrm{km}$ ocean area, where the horizontal coordinates satisfy $(x, y) \in [-750, 750]\,\mathrm{m}$.
The system consists of \(D=3\) HAPS-BSs operating at an altitude \(h=20\)\,km, 
with a carrier frequency of $f=3.5\,\mathrm{GHz}$ and total bandwidth $B=100\,\mathrm{MHz}$. 
Each HAPS-BS transmits with a power $P^{tx}=55$\,dBm, and a noise power is \SI{-114}{dBm/MHz}. 
%and employs an antenna with a $3$~dB beamwidth of $65^\circ$, element gain of 8~dBi, and front-to-back ratio of 30~dB.  %%[German]: these are UPA antenna parameters, not aplicable here. In general, we described reflector antenna in the System model with a reference to corresponding standard.
We consider three hotspots, representing ship clusters, each with a radius of $50\,\mathrm{m}$ and a density of 10 UEs, resulting in a total of $U = 30$ UEs located at an altitude of $h_{\rm UE} = 1.5\,\mathrm{m}$.
We consider a linear motion model in which UE hotspots originate from the initial positions
\[
(-550\,\text{m}, -550\,\text{m}), \quad (550\,\text{m}, 0), \quad (-550\,\text{m}, 550\,\text{m}).
\]
Each hotspot moves along a straight-line trajectory with a constant velocity of $10\,\text{m/s}$, representing coordinated group mobility within predefined spatial boundaries.
%The propagation environment follows the 3GPP TR 38.811 UMa model with a Rician channel, and a noise figure of 5~dB. %%[German]: propagation model, basically, described in the System model section
% We implement a linear motion model, where UE hotspots (ships) move in a straight line at a constant speed of 10~m/s, 
% representing coordinated group movement between the specified boundaries. 
% % The direction of motion is from right to left, and boundary reflections are applied when the hotspots reach the area limits.

\subsubsection*{PPO Algorithm Configuration}
The proposed PPO-based algorithm is trained over 12k episodes, 
each consisting of 128 frames. 
The model uses a learning rate of $3 \times 10^{-5}$, 
with three hidden layers (128 neurons each) for both actor and critic networks. 
A discount factor $\gamma^{\mathrm{df}}=0.99$ %% [German]: gamma is already used for SINR, so this may be confusing with the following formula (17). I added ^{df} but maybe you find a better notation
and generalized advantage estimation parameter $\lambda=0.95$ are used to balance short- and long-term rewards. 
The PPO clip parameter is set to $\epsilon=0.2$, with an entropy coefficient of 0.1 to encourage exploration. 
Gradient clipping with a maximum norm of 1.0 is applied for training stability. A state memory size of $M=1$ is used, enabling the agent to capture UE mobility trends without direct position information. For the state space, assuming three UAV-BSs operating at a fixed altitude and serving three hotspots, the overall state space is defined as:
% --- Reduce equation spacing locally ---
\begingroup
\setlength{\abovedisplayskip}{6pt}
\setlength{\belowdisplayskip}{6pt}

\begin{equation}
\label{eq:observation_set}
\mathcal{\mathbf{S}} =
\left\{
\begin{aligned}
&\big((x_{1,i},y_{1,i},z_{1,i}=20\,\mathrm{km}),\, \bar{\gamma}_{1,1,i},\, \Psi_{1,1,i},\, \Phi_{1,1,i}\big),\\
&\big((x_{2,i},y_{2,i},z_{2,i}=20\,\mathrm{km}),\, \bar{\gamma}_{2,2,i},\, \Psi_{2,2,i},\, \Phi_{2,2,i}\big),\\
&\big((x_{3,i},y_{3,i},z_{3,i}=20\,\mathrm{km}),\, \bar{\gamma}_{3,3,i},\, \Psi_{3,3,i},\, \Phi_{3,3,i}\big).
\end{aligned}
\right\}.
\end{equation}

\endgroup
For the reward $\mathcal{R} = \left(1 + \exp\!\left(-c_s (R_{\mathrm{fair}} - c_m)\right)\right)^{-1}$, we set $c_s = 0.25$ and $c_m = 50$, based on the throughput achieved when HAPS-BSs are directly above their corresponding hotspots and the average throughput over entire episodes across multiple realizations, providing a stable reference for reward scaling.
\subsubsection*{Wind Model Configuration}

The stratospheric wind model is configured with the following parameters:
$\Delta t = 2$ s, spatial scale $= 1000$ m/unit, mean wind speed $= 4$ m/s, residual standard deviation $= 2$ m/s, temporal correlation coefficient $\rho^{\mathrm{temp}} = 0.95$, %% [German]: same, as above, rho is already used so I added ^{temp} but feel free to change
slow variation amplitude $= 1$ m/s, and slow variation period $= 128$ steps.
\begin{table}[b]
\centering
\caption{Predefined HAPS-BS deployment scenarios}
\label{tab:uav_scenarios}
\begin{tabular}{c c}
\hline
\textbf{Scenario} & \textbf{Horizontal Coordinates $(x,y)$} \\
\hline
 & $\text{HAPS}-\text{BS}_{1}, \text{HAPS}-\text{BS}_{2}, \text{HAPS}-\text{BS}_{3}$ \\
\hline
1 & $[(-250, -450), (450, 0), (-250, 450)]$ \\
2 & $[(-450, 450), (-100, 0), (-450, -450)]$ \\
3 & $[(-450,200), (100, 100), (-450, -200)]$ \\
4 & $[(0, 0), (0, 0), (0, 0)]$ \\
\hline
\end{tabular}
\end{table}
\subsubsection*{Results and Discussion}
To evaluate generalization, the proposed system is assessed during both training and evaluation phases under stratospheric wind disturbances. During training, the initial HAPS-BS positions are randomly initialized, and the corresponding reward and throughput are monitored (see Fig.~2(a) and Fig.~2(b)). During evaluation, the HAPS-BSs are initialized at a fixed altitude of $20\,\mathrm{km}$ under predefined deployment scenarios, with horizontal coordinates summarized in Table~I. The performance is evaluated every 500 iterations across all scenarios.
As shown in Fig.~2(a) and Fig.~2(b), both the training reward and throughput exhibit a consistent upward trend despite the presence of wind-induced perturbations, indicating that the agent progressively learns a robust positioning policy. While fluctuations are observed during early stages due to stochastic wind dynamics and exploration, the curves gradually stabilize, suggesting convergence under disturbed conditions.
The evaluation results in Fig.~2(c) and Fig.~2(d) further demonstrate that the learned policy generalizes well across different initial configurations in the presence of wind variability. Despite the diversity of the four scenarios and environmental disturbances, the performance remains stable and converges to comparable levels in terms of both reward and throughput. Minor variations observed at early iterations are primarily due to initial positioning offsets and wind-induced drift.
The results show that the proposed DRL-based approach compensates for wind-induced deviations and maintains stable performance across varying initial deployments in dynamic maritime environments.

\section{Conclusion}
%{\color{red}
This paper proposes a DRL-based approach for dynamic HAPS-BS positioning in maritime networks. Leveraging PPO and realistic radio measurements, it enables adaptive positioning in response to UE mobility, channel variations, and wind-induced perturbations without requiring explicit location information. The framework models continuous HAPS-BS mobility under stratospheric wind dynamics and supports efficient control in dynamic environments. Simulation results show stable convergence and consistent performance across scenarios, demonstrating good generalization. Overall, the method achieves reliable and stable network performance with sufficient throughput, highlighting the potential of intelligent HAPS-BS systems for wide-area maritime connectivity.
%}
%As a future work, we will consider deploying multiple UAVs as aerial BSs to further enhance network coverage and reliability. 
%As the future work, we plan to explore advanced coordination strategies and dynamic resource allocation among the UAVs. Additionally, we aim to integrate more complex UE mobility models and environmental factors to improve the robustness and scalability of our approach.}

\vspace{12pt}

\bibliography{bib}

\end{document}